\documentclass[aps,prl,twocolumn,showpacs,superscriptaddress]{revtex4}

\usepackage{graphicx}  % needed for figures
\usepackage{dcolumn}   % needed for some tables
\usepackage{bm}        % for math
\usepackage{amssymb}   % for math
\usepackage{amsmath}
\usepackage{units}
\usepackage[dvipsnames]{xcolor}
\usepackage[normalem]{ulem} % stricken out text for edits

\usepackage[type1]{libertine}
\usepackage{textcomp}% Required to get special symbols
\usepackage[scaled=.85]{beramono}% Typewriter font
\usepackage[libertine,cmintegrals,cmbraces,vvarbb,slantedGreek]{newtxmath}
\usepackage[scr=boondoxo]{mathalfa}% Extra math symbols
\usepackage{bm}% Extra bold faces
\usepackage[lf]{carlito}

\newcommand{\vek}[1]{\boldsymbol{#1}}
\newcommand{\cev}[1]{\reflectbox{\ensuremath{\vec{\reflectbox{\ensuremath{#1}}}}}}

\makeatletter
\DeclareRobustCommand{\cev}[1]{%
  \mathpalette\do@cev{#1}%
}
\newcommand{\do@cev}[2]{%\overleftarrow{}
  \fix@cev{#1}{+}%
  \reflectbox{$\m@th#1\vec{\reflectbox{$\fix@cev{#1}{-}\m@th#1#2\fix@cev{#1}{+}$}}$}%
  \fix@cev{#1}{-}%
}
\newcommand{\fix@cev}[2]{%
  \ifx#1\displaystyle
    \mkern#23mu
  \else
    \ifx#1\textstyle
      \mkern#23mu
    \else
      \ifx#1\scriptstyle
        \mkern#22mu
      \else
        \mkern#22mu
      \fi
    \fi
  \fi
}
\makeatother

\renewcommand{\Re}{\text{Re}}
\renewcommand{\Im}{\text{Im}}

\newenvironment{alexdes}
    {\begin{list}{}%
        {%
         \setlength{\labelwidth}{0pt}%
         \setlength{\labelsep}{0pt}%
         \setlength{\leftmargin}{0pt}
         \setlength{\topsep}{0mm}
         \setlength{\itemsep}{0.5mm}
         \setlength{\parsep}{0mm}
        }%
    }%
    {\end{list}}

\begin{document}

\title{Robustness of Yu-Shiba-Rusinov resonances in presence of a complex superconducting order parameter}

\author{Jacob Senkpiel}
\affiliation{Max-Planck-Institut f\"ur Festk\"orperforschung, Heisenbergstraße 1,
70569 Stuttgart, Germany}
\author{Carmen Rubio-Verd{\'u}}
\affiliation{Max-Planck-Institut f\"ur Festk\"orperforschung, Heisenbergstraße 1,
70569 Stuttgart, Germany}
\affiliation{CIC nanoGUNE, 20018 Donostia-San Sebasti\'an, Spain}
\author{Markus Etzkorn}
\affiliation{Max-Planck-Institut f\"ur Festk\"orperforschung, Heisenbergstraße 1,
70569 Stuttgart, Germany}
\author{Robert Drost}
\affiliation{Max-Planck-Institut f\"ur Festk\"orperforschung, Heisenbergstraße 1,
70569 Stuttgart, Germany}
\author{Leslie M. Schoop}
\affiliation{Department of Chemistry, Princeton University, Princeton, NJ 08544, USA}
\author{Simon Dambach}
\affiliation{Institut für Komplexe Quantensysteme and IQST, Universität Ulm, Albert-Einstein-Allee 11, 89069 Ulm, Germany}
\author{Ciprian Padurariu}
\affiliation{Institut für Komplexe Quantensysteme and IQST, Universität Ulm, Albert-Einstein-Allee 11, 89069 Ulm, Germany}
\author{Bj\"orn Kubala}
\affiliation{Institut für Komplexe Quantensysteme and IQST, Universität Ulm, Albert-Einstein-Allee 11, 89069 Ulm, Germany}
\author{Joachim Ankerhold}
\affiliation{Institut für Komplexe Quantensysteme and IQST, Universität Ulm, Albert-Einstein-Allee 11, 89069 Ulm, Germany}
\author{Christian R. Ast}
\email[Corresponding author; electronic address:\ ]{c.ast@fkf.mpg.de}
\affiliation{Max-Planck-Institut f\"ur Festk\"orperforschung, Heisenbergstraße 1,
70569 Stuttgart, Germany}
\author{Klaus Kern}
\affiliation{Max-Planck-Institut f\"ur Festk\"orperforschung, Heisenbergstraße 1,
70569 Stuttgart, Germany}
\affiliation{Institut de Physique, Ecole Polytechnique Fédérale de Lausanne, 1015 Lausanne, Switzerland}

\date{\today}

\begin{abstract}
Robust quantum systems rely on having a protective environment with minimized relaxation channels. Superconducting gaps play an important role in the design of such environments. The interaction of localized single spins with a conventional superconductor generally leads to intrinsically extremely narrow Yu-Shiba-Rusinov (YSR) resonances protected inside the superconducting gap. However, this may not apply to superconductors with more complex, energy dependent order parameters. Exploiting the Fe-doped two-band superconductor NbSe$_2$, we show that due to the nontrivial relation between its complex valued and energy dependent order parameters, YSR states are no longer restricted to be inside the gap. They can appear outside the gap (i.\ e.\ inside the coherence peaks), where they can also acquire a substantial intrinsic lifetime broadening. \textit{T}-matrix scattering calculations show excellent agreement with the experimental data and relate the intrinsic YSR state broadening to the imaginary part of the host's order parameters. Our results suggest that non-thermal relaxation mechanisms contribute to the finite lifetime of the YSR states, even within the superconducting gap, making them less protected against residual interactions than previously assumed. YSR states may serve as valuable probes for nontrivial order parameters promoting a judicious selection of protective superconductors.
\end{abstract}

\pacs{74.55.+v, 74.25.-q, 74.25.F-, 74.45.+c}

\maketitle

\subsection{Introduction}

For the past decades, YSR states \cite{yu_bound_1965,shiba_classical_1968,rusinov_superconductivity_1969} have been used as local probes to study superconductivity \cite{yazdani_probing_1997}, adsorbate-substrate interaction \cite{hudson_interplay_2001}, their interplay with the Kondo effect \cite{franke_competition_2011} as well as the properties of the impurity spin states themselves \cite{ruby_orbital_2016}. The interest in YSR states has intensified in recent years as they play a vital role in engineering Majorana bound states \cite{nadj-perge_proposal_2013,nadj-perge_observation_2014,ruby_end_2015} as well as in studying topological superconductors \cite{menard_two-dimensional_2017}. In a simple Bardeen-Cooper-Schrieffer (BCS)-type $s$-wave superconductor with a real-valued energy-independent order parameter \cite{balatsky_impurity-induced_2006}, YSR states can only exist inside the superconducting gap, which protects them from interacting with, and decaying into, the quasiparticle continuum \cite{heinrich_protection_2013}. If this gap is void of quasiparticles, only a thermally induced decay into the quasiparticle continuum is possible \cite{ruby_tunneling_2015}. Residual quasiparticle interactions could slightly broaden the YSR state \cite{dynes_direct_1978,martin_nonequilibrium_2014}. The situation is different in a $d$-wave superconductor, where the YSR state is intrinsically expected to have a non-zero lifetime broadening inside the gap owing to its nontrivial order parameter \cite{salkola_spectral_1997,hudson_interplay_2001}. However, as we will show here, the behavior of YSR states changes dramatically even for $s$-wave superconductors if they feature a nontrivial order parameter.

We choose an $s$-wave two-band superconductor with finite interband coupling resulting in complex-valued and energy-dependent order parameters \cite{suhl_bardeen-cooper-schrieffer_1959,mcmillan_tunneling_1968}. This leads not only to a nontrivial relation between the order parameter and the position of the gap edge, but, due to causality \cite{toll_causality_1956}, also implies imaginary parts in the order parameters, which cannot be trivially removed by a gauge transformation.  Effectively, intrinsic decay channels for YSR states emerge. Hence, the interplay between interband coupling and YSR states reveals fundamental properties of superconductors that may be of relevance for other unconventional materials.

We explore YSR states in the Fe-doped two-band superconductor NbSe$_2$, which has been well studied in the past \cite{hess_vortex-core_1990,hayashi_effects_1997,yokoya_fermi_2001,rodrigo_stm_2004,kohen_probing_2006,guillamon_intrinsic_2008,kimura_josephson_2009,noat_signatures_2010,rahn_gaps_2012,menard_coherent_2015,noat_quasiparticle_2015,ptok_yu-shiba-rusinov_2017,galvis_tilted_2017,dvir_spectroscopy_2018}. Due to interband coupling, a BCS-type band induces superconductivity also in a second band (proximity effect), so that the individual order parameters turn out to be strongly energy dependent. The Fe-doping gives rise to YSR states that can be directly observed with STM for impurities near the surface. In this way, we demonstrate that in this two-band superconductor, the energy of YSR states are no longer restricted to inside the gap, but are also found within the quasi-particle continuum. More specifically, those YSR states with stronger magnetic exchange coupling are located within the superconducting gap and have a very small intrinsic lifetime broadening due to a reduced relaxation. We can also safely neglect thermally activated relaxation processes for YSR states within the gap \cite{ruby_tunneling_2015}, as we operate at a base temperature of 15\,mK, which is more than two orders of magnitude below the superconducting transition temperature. For a weaker exchange coupling, the YSR states are located outside of the superconducting gap within the coherence peaks, where they broaden substantially with strong intrinsic relaxation channels to the superconducting host. We demonstrate that the enhanced lifetime broadening is directly related to the proximity-induced complex order parameters. Their imaginary parts are associated with relaxation processes within the superconductor. The interaction of an impurity with the individual bands in the bulk makes these decay channels available to the YSR states. In this way, we do not only demonstrate the relevance of intrinsic relaxation channels for a certain class of robust quantum states, but also establish YSR states as a probe for the imaginary part of a complex-valued and energy dependent order parameter.

\begin{figure}
\centerline{\includegraphics[width = \columnwidth]{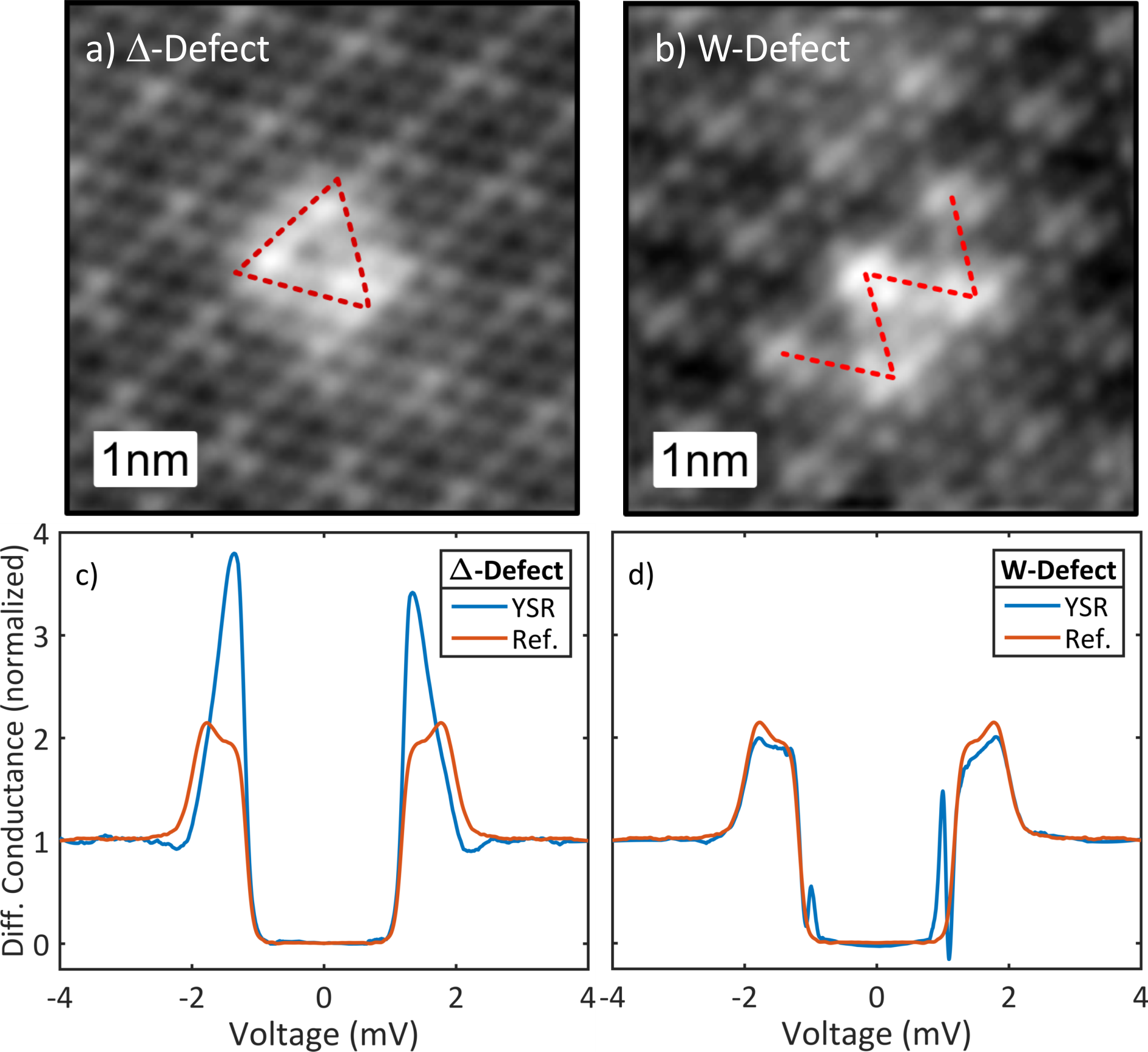}}
\caption{\textbf{Topography and differential conductance spectra of Fe in NbSe$_2$:} Topographies of the two most common Fe impurities having a triangular shape ($\upDelta$) (a) and a W-shape (W) (b). The triangular defect typically has a smaller exchange coupling so that the YSR states appear within the coherence peaks (c). The W-defect has typically a higher exchange coupling and the YSR states commonly appear inside the gap. An unperturbed reference spectrum with no Fe impurity in the vicinity is shown in red. The current setpoint for the topography was 20\,pA at a bias voltage of 100\,mV, the setpoint for the spectra was 200\,pA at 4\,mV.} \label{fig:spectra}
\end{figure}

\subsection{Experimental}

Single crystals of Fe-doped NbSe$_2$ were grown by chemical vapor transport. Powders of Nb and Fe in a ratio of 99.45:0.55 were mixed well and then placed in a quartz tube. Se chips were added in a stoichiometric ratio to yield Nb$_{0.9945}$Fe$_{0.055}$Se$_2$. Iodine was used as the transport agent. The sealed tube was heated to 900\,$^\circ$C with a temperature gradient of 75\,$^\circ$C for three weeks. The structure and composition of the crystals was confirmed with X-ray diffraction.

Experiments were carried out in a home built STM operating at 15\,mK base temperature. Samples were cleaved using scotch tape in ultra-high vacuum. To enhance the experimental resolution, superconducting vanadium was chosen as a tip material \cite{franke_competition_2011}. Tips were cut from 0.1\,mm vanadium wire (Goodfellow) in air and cleaned by Ar$^+$ ion sputtering in UHV followed by field emission on a V(100) sample. Tips were characterised on a clean V(100) surface before measurements on NbSe$_2$ and the tip gap extracted by fitting SIS tunnel spectra while assuming the bulk gap of the vanadium sample to be $\Delta_{\mathrm{V}} = 710\,\upmu e$V.

\subsection{Characterizing Fe-doped NbSe$_2$}

Layered NbSe$_2$ is a two-band superconductor, whose bands interact via electron hopping between states near the Fermi edge \cite{suhl_bardeen-cooper-schrieffer_1959,mcmillan_tunneling_1968,machida_bound_1972,yokoya_fermi_2001,rodrigo_stm_2004,noat_signatures_2010,rahn_gaps_2012,noat_quasiparticle_2015,dvir_spectroscopy_2018}. We follow the description by McMillan to model this mechanism \cite{mcmillan_tunneling_1968}. Without the interband coupling, the first band is commonly assumed to be superconducting, while the second is not \cite{noat_signatures_2010}. Only the interband coupling induces superconductivity in the second band, which in turn reduces the order parameter in the first band. As a result, two energy dependent order parameters emerge, which are complex-valued due to causality \cite{toll_causality_1956}. The imaginary part can be interpreted as an intrinsic inverse lifetime due to the hopping between the bands. This stands in contrast to the real-valued energy independent BCS-type order parameter.

In order to induce YSR states in the NbSe$_2$ host, we have doped the crystal with about 0.55\% Fe \cite{supinf}. Dopants that are close to the surface can be directly seen in the topographic image shown in Fig.\ \ref{fig:spectra}. We find two characteristic types of impurities in our samples, which we attribute to Fe-defects: one with a triangular ($\upDelta$) shape (Fig.\ \ref{fig:spectra}a) and one with a W-shape (Fig.\ \ref{fig:spectra}b). Both give rise to strong YSR states as can be seen in the differential conductance spectra measured with a superconducting vanadium tip in Fig.\ \ref{fig:spectra} c) and d), respectively. However, the $\upDelta$-defect shows the YSR state inside the coherence peaks (Fig.\ \ref{fig:spectra} c)), as can be seen by comparison with the spectrum on the bare surface (red line). Indeed, the asymmetry of the peak heights suggests the existence of YSR states as opposed to a change of the local tunneling probability into the two different bands. By contrast, the YSR state in Fig.\ \ref{fig:spectra}d appears close to the gap edge, but clearly inside the gap. Evidently, these two types of YSR states substantially differ in line width. The in-gap line width (Fig.\ \ref{fig:spectra}d)) is much narrower than the line width outside the gap (Fig.\ \ref{fig:spectra}c)). These experimental results present quite a different appearance of YSR states than the ``conventional'' extremely narrow features occurring only inside the superconducting gap \cite{heinrich_single_2017}. In the following, we will demonstrate that this is a direct consequence of the energy dependent order parameter.

Both topographic images in Fig.\ \ref{fig:spectra} show a rather regular, continuous lattice corrugation only modulated by a stronger density of states at the defect positions suggesting that the impurities are buried very close to the surface, but not directly at or on the surface. For further analysis, we subtract an unperturbed reference spectrum of the bare surface, i.\ e.\ with no impurities in the close vicinity of the YSR state, in order to isolate the YSR states.

\subsection{The Order Parameters in Fe-doped NbSe$_2$}

In order to find a simple yet appropriate theoretical model, we need a detailed description of the order parameter in Fe-doped NbSe$_2$. The doping of 0.55\% of Fe atoms is already strong enough to reduce the transition temperature from about 7.2\,K \cite{frindt_superconductivity_1972,revolinsky_superconductivity_1965,yokoya_fermi_2001} to 6.1\,K  \cite{supinf}, so that the effects of the magnetic impurities on the bulk superconductor cannot be neglected. In order to theoretically describe the superconducting order parameter, we have to include the interaction between the two bands \cite{suhl_bardeen-cooper-schrieffer_1959,mcmillan_tunneling_1968, schopohl_tunneling_1977}, as well as the interaction with a small but finite concentration of magnetic impurities \cite{abrikosov_conribution_1960, maki_pauli_1964, shiba_classical_1968}. As we are analyzing density of states measurements with no momentum information on the band structure nor on the order parameter, we refrain from employing models involving momentum dependent order parameters \cite{hayashi_effects_1997,rahn_gaps_2012,galvis_tilted_2017}, but instead focus on an effective description of the order parameter (for details, see the Supporting Information \cite{supinf}). We add the selfenergies of the two interactions to the bare order parameters $\Delta^\text{BCS}_{1,2}$ for the first and second band, respectively and find two coupled equations for the two order parameters $\Delta_{1,2}(\omega)$ \cite{zarate_phonon_1985,golubov_effect_1997}:

\begin{eqnarray}
    \label{eq:delta}
    \Delta_1(\omega) &=& \Delta_{1}^\text{BCS} - \Gamma_{12}\frac{\Delta_1(\omega)-\Delta_2(\omega)}{\sqrt{\Delta^2_2(\omega) - \omega^2}}- \zeta_1\frac{\Delta_1(\omega)}{\sqrt{\Delta^2_1(\omega) - \omega^2}}\nonumber\\[-5pt]
\\[-5pt]
    \Delta_2(\omega) &=& \Delta_{2}^\text{BCS} - \Gamma_{21}\frac{\Delta_2(\omega)-\Delta_1(\omega)}{\sqrt{\Delta^2_1(\omega) - \omega^2}}- \zeta_2\frac{\Delta_2(\omega)}{\sqrt{\Delta^2_2(\omega) - \omega^2}}\nonumber.
\end{eqnarray}

Here, $\omega$ is the energy, $\Gamma_{12}$ and $\Gamma_{21}$ are the coupling parameters between the two bands and $\zeta_{1,2}$ are the coupling constants of the first and second band to the finite concentration of magnetic impurities. The interband hopping as well as the impurity interaction are proportional to the density of states $n_{1,2}$ at the Fermi level of each band, so that the parameters in Eq.\ \ref{eq:delta} are related in the following way:

\begin{equation}
    \frac{\Gamma_{21}}{\Gamma_{12}} = \frac{\zeta_1}{\zeta_2} = \frac{n_1}{n_2}.
\end{equation}

Eq.\ \ref{eq:delta} can be solved numerically using a multi-dimensional Newton-Raphson method. A more detailed discussion of these equations is given in the Supporting Information \cite{supinf}.

The resulting normalized density of states $\rho_i(\omega)$ of the $i$-th band is

\begin{equation}
    \rho_i(\omega) = \Re\left[\frac{\omega}{\sqrt{\omega^2-\Delta_i^2(\omega)}}\right],
\end{equation}

so that the total normalized density of states can be written as:

\begin{equation}
    \rho(\omega) = \frac{1+\eta}{2}\rho_1(\omega) + \frac{1-\eta}{2}\rho_2(\omega),\label{eq:rho}
\end{equation}

where $\eta$ is a ratio that accounts for the different densities of states as well as the different tunneling probabilities into the two bands. Using the total density of states convolved with the superconducting density of states of the V tip and with the energy resolution function \cite{ast_sensing_2016}, we can fit the model to our experimental data. For details, see the Supporting Information \cite{supinf}.

\begin{figure}
\centerline{\includegraphics[width = 0.95\columnwidth]{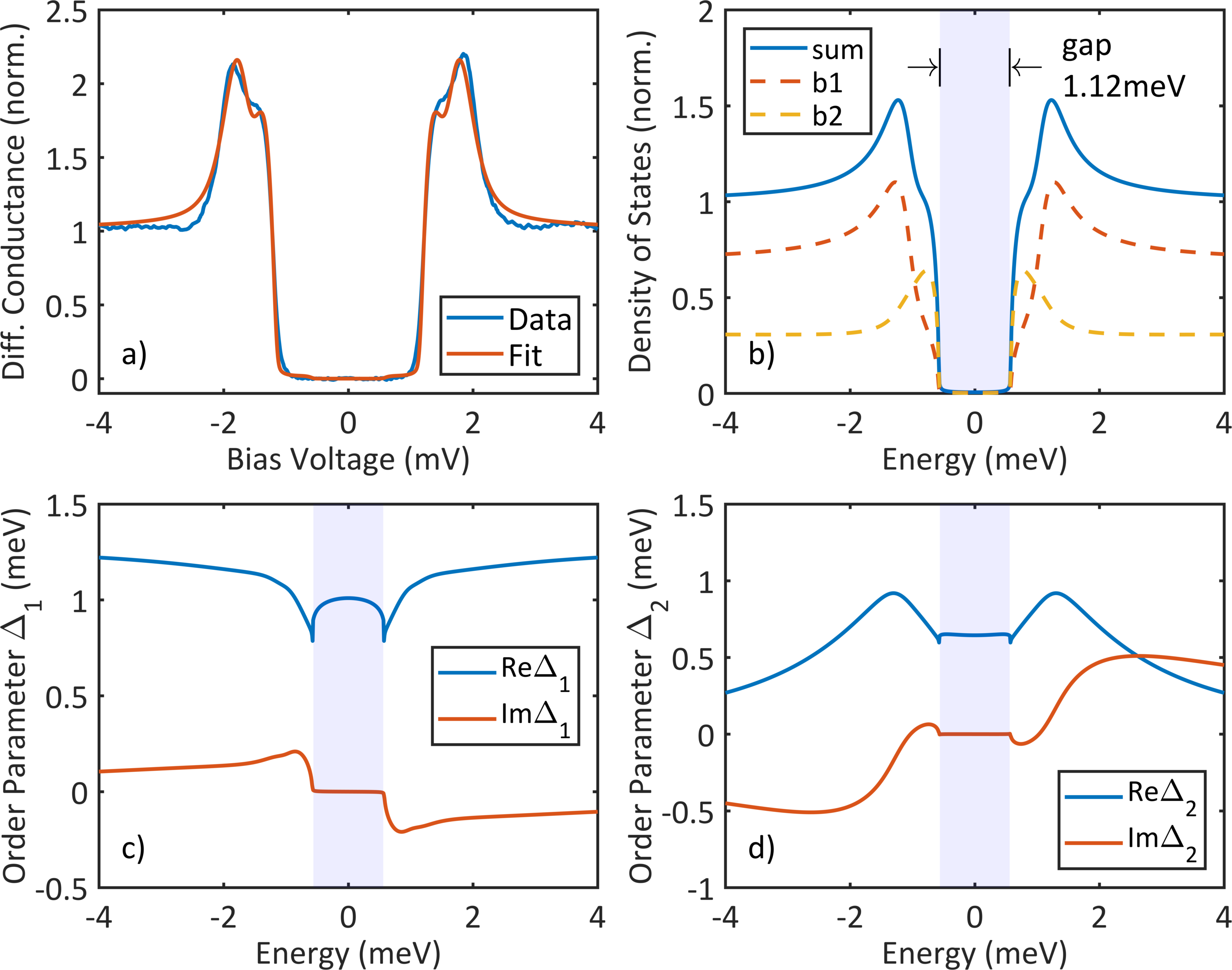}}
\caption{\textbf{Extracting the order parameters of Fe-doped NbSe$_2$:} (a) Fit of the interband-impurity model to an unperturbed conductance spectrum. The extracted fit parameters provide the input values for the subsequent analysis. (b) Calculated total density of states of the superconducting substrate from the extracted fit parameters (sum). The weighted density of states for band 1 (b1) and band 2 (b2) are shown as dashed lines. Note that the gap edges of both bands are at the same energy. (c) Resulting order parameter $\Delta_1(\omega)$ of the first band. (d) Resulting order parameter $\Delta_2(\omega)$ of the second band. The blue shaded region indicates the gap.} \label{fig:fitsops}
\end{figure}

The resulting fit captures the experimental differential conductance quite accurately as shown in Fig.\ \ref{fig:fitsops}a). Following a previous analysis \cite{noat_signatures_2010}, we have assumed a second band that is intrinsically normal conducting, $\Delta_{2}^\text{BCS}=0$. The best fits are obtained for a density of states ratio of $n_1/n_2=5$, in agreement with previous assessments \cite{noat_signatures_2010}. The unperturbed order parameter for the first band $\Delta_{1}^\text{BCS}=1.27\,$meV is somewhat smaller than what has been reported for undoped NbSe$_2$ ($\Delta_{1}^\text{BCS,lit}=1.4\,$meV, see Ref.\ \onlinecite{noat_signatures_2010} and references therein), but corresponds to roughly the same ratio as the reduction in the transition temperature from 7.2\,K to 6.1\,K. For the coupling terms, we find $\Gamma_{12}=0.36$\,meV, $\zeta_1=57\,\upmu$eV, and $\eta = 0.38$. The extracted density of states of Fe-doped NbSe$_2$ is plotted in Fig.\ \ref{fig:fitsops}b). The weighted individual densities of states are shown as dashed lines. Their gap edges lie at the same energy. The total density of states (solid line) features rather blunt coherence peaks, where shoulders indicate the coherence peaks of the second band. The gap itself is substantially narrower (1.12\,meV) than the bare BCS gap ($2\Delta_{1}^\text{BCS}=2.54$\,meV) (see Supporting Information \cite{supinf}). The corresponding order parameters are plotted in Fig.\ \ref{fig:fitsops}c) and d). For large energies the real part of the order parameter for the first band shows an asymptotic approach to the $\Delta_{1}^\text{BCS}$ value of 1.27\,meV, while the imaginary part approaches zero. Inside the gap, the order parameter is entirely real valued. In the vicinity of the coherence peaks, however, a strong energy dependence is visible. The order parameter of the second band approaches zero for large energies.

As a consequence, it becomes clear that for energies below and around the coherence peaks, the real parts of the order parameters shown in Fig.\ \ref{fig:fitsops}c) and d) are larger than the energy of the gap edge ($\pm0.55\,$meV). Because the energy position of the YSR state is directly related to the value of the order parameter, we can already anticipate unconventional locations of the YSR states \cite{salkola_spectral_1997,flatte_local_1997}.

\subsection{Magnetic impurity scattering of Fe in NbSe$_2$}

We now turn to the impurity scattering and calculate the YSR spectra following a simple T-matrix scattering formalism \cite{salkola_spectral_1997}. We follow a mean field approach in assuming that the $\zeta_{1/2}$ terms affect the sample DOS on a global scale, while YSR scattering is local. The superconducting host is described by the normalized Green's function for the two bands $G_{1,2}(\omega)$ using the energy-dependent order parameters discussed in the previous section:

\begin{eqnarray}
G_{1,2}(\omega)&=&-\pi\frac{1\pm\eta}{2}\frac{(\omega +  i\Gamma)\sigma_0 - \Delta_{1,2}(\omega)\sigma_1}{\sqrt{\Delta_{1,2}^2(\omega) - (\omega +  i\Gamma)^2}}. \label{eq:G0}
\end{eqnarray}

Here, the $\sigma_i$ are the Pauli matrices in Nambu space with $\sigma_0$ being the identity matrix. We add a Dynes-type parameter $\Gamma \leq 5\,\upmu$eV as a phenomenological lifetime broadening \cite{dynes_direct_1978}. Larger values of $\Gamma$ fill the gap, which is not observed in the experiment. The T-matrix for the $i$th band can be written as

\begin{equation}
 T_i(\omega) = V_i(1 - G_i(\omega)V_i)^{-1}\ \text{with}\  V_i = J'_i\sigma_0 + U'_i\sigma_3,
\end{equation}

where $V_i$ is the scattering potential with $J'_i=\frac{1}{2}JSn_i$ as the dimensionless, effective magnetic exchange coupling and $U'_i = Un_i$ as the dimensionless, effective local Coulomb scattering. Furthermore, $\frac{1}{2}JS$ is the exchange coupling of a classical spin, $U$ is the local Coulomb potential, and $n_i$ is the density of states of the $i$th band at the Fermi level. The total Green's function $G_i^\text{YSR}(\omega)$ can be written as:

\begin{equation}
 G_i^\text{YSR}(\omega) = G_i(\omega) + G_i(\omega)T_i(\omega)G_i(\omega).\label{eq:ysr}
\end{equation}

For simplicity we only consider the total Green's function at the position of the impurity. Inserting the energy dependent order parameters into Eq.\ \ref{eq:G0}, we can calculate the spectral function $A_i(\omega)=-\frac{1}{\pi}\text{tr}\{\Im[G_i^\text{YSR}(\omega)]\}$ and thus the density of states of the YSR resonances for different exchange couplings $J'_i$.

\begin{figure}
\centerline{\includegraphics[width = \columnwidth]{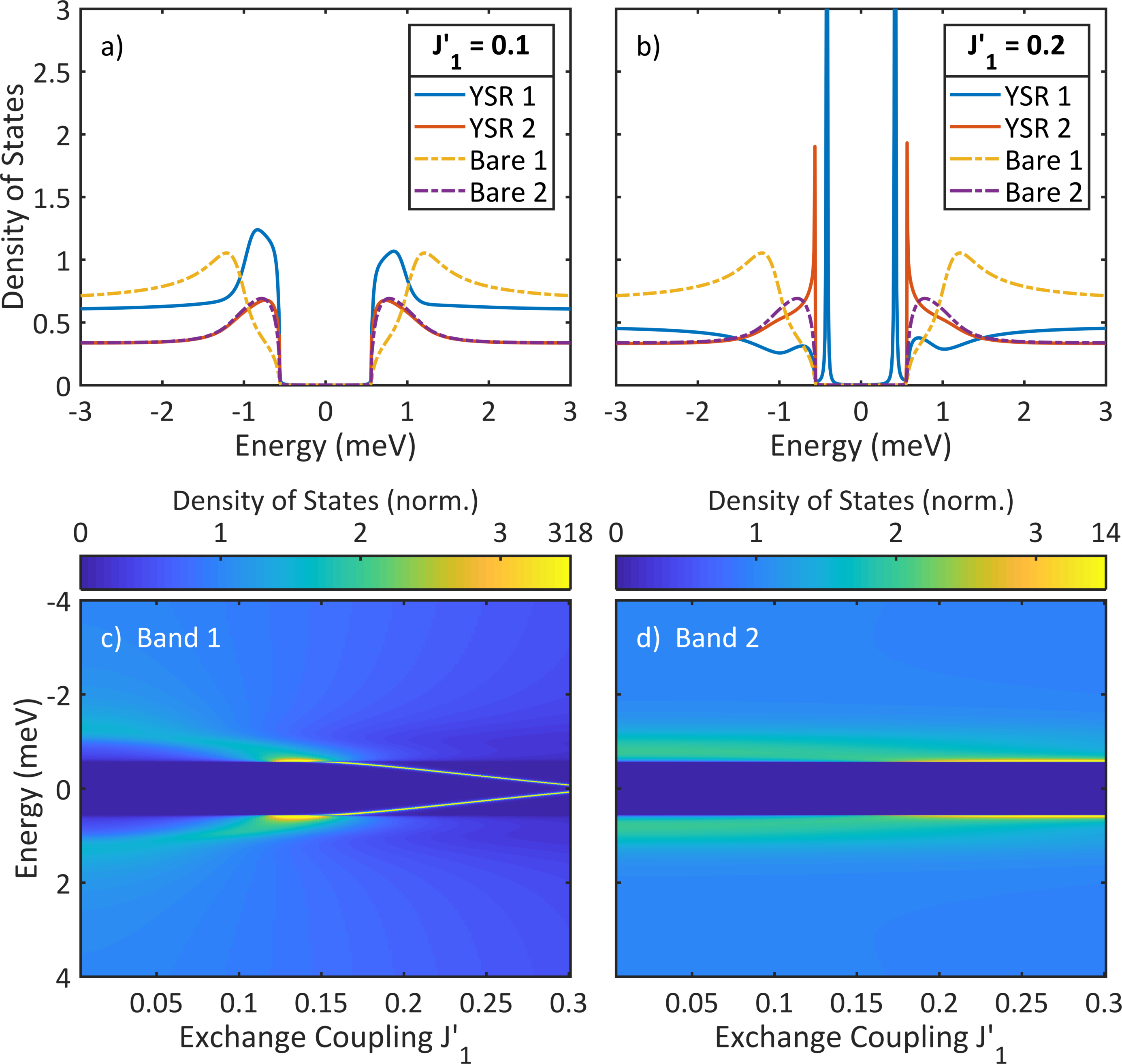}}
\caption{\textbf{Calculated YSR states:} The solid lines correspond to the YSR spectra coupling to band 1 (YSR 1) and band 2 (YSR 2), while the dashed lines represent the unperturbed spectra for the first (Bare 1) and second band (Bare 2), respectively. The ratio between the spectra of the two bands correspond to the typical ratio $\eta$ observed in the experiment. A small value for the Coulomb interaction $U'_1=0.05$ and $U'_2=0.01=U'_1n_2/n_1$ has been chosen to make the spectrum asymmetric. (a) YSR spectrum for weak exchange coupling. The YSR state is broad and within the region of the coherence peaks. (b) Stronger coupling than (a), but still before the zero bias crossing. The YSR state resides within the gap and has become extremely sharp. (c) and (d) YSR spectra vs.\ exchange coupling in band 1 and band 2, respectively. The color scaling is adapted to show the weaker features; it is non-linear for values larger than 3.5. The sharpening of the peak as it enters the gap is clearly visible. In (d) the YSR peak develops much slower due to the reduced density of states in band 2. For all panels, $J'_1/J'_2=n_1/n_2$.} \label{fig:shibacalc}
\end{figure}

The calculated YSR resonances for different exchange couplings $J'_1=(n_1/n_2)J'_2$ are plotted in Fig.\ \ref{fig:shibacalc}. For weak exchange coupling, the YSR state interacting with the first band can be found within the coherence peaks (cf.\ Fig.\ \ref{fig:shibacalc}a)). For stronger exchange coupling (but still below the zero energy crossing \cite{franke_competition_2011}), the YSR state for the first band moves into the gap and becomes extremely sharp, as can be seen in Fig.\ \ref{fig:shibacalc}b). Inside the gap, the width of the YSR state is given by the parameter $\Gamma$. In Fig.\ \ref{fig:shibacalc}c) and d), the YSR spectra are shown as a function of exchange coupling $J'$ for the first and second band, respectively. With increasing coupling to the first band, the YSR peaks shift towards zero energy and become extremely narrow when entering the gap region.

Apparently, the second band affects the YSR states much less, as can be seen in Figs.\ \ref{fig:shibacalc}c) and d). Both bands have predominantly Nb-$d$ character \cite{rahn_gaps_2012,noat_quasiparticle_2015}. However, as the effective exchange coupling between the impurity and the superconductor is proportional to the density of states in the superconductor, i.\ e.\ $J'_1/J'_2=n_1/n_2$, we may assume a strongly reduced impact of the second band. We therefore consider in leading order only the YSR resonances in the first band.

\begin{figure}
\centerline{\includegraphics[width = \columnwidth]{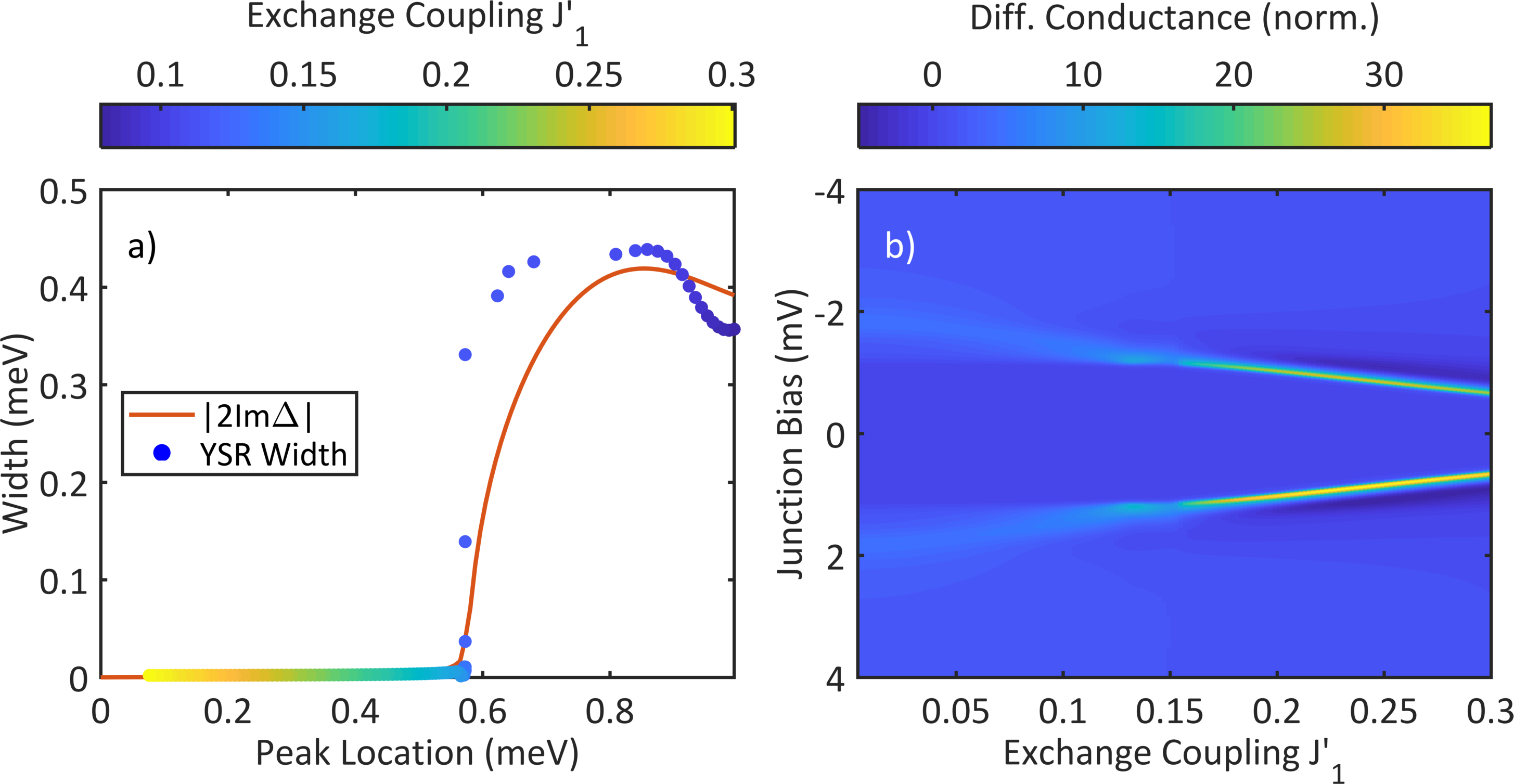}}
\caption{\textbf{Evolution of the YSR peak width:} (a) YSR peak width vs.\ peak location. The width becomes extremely narrow in the region of the superconducting gap. The color coding represents the strength of the exchange coupling. The red line is twice the imaginary part of the order parameter $\Delta_1(\omega)$ of the first band. We see a clear correlation between the peak width and the imaginary part. (b) Differential conductance calculated from the YSR spectral function and the density of states of the superconducting V tip as well as finite energy resolution. The sharpening of the YSR state when entering the gap, is clearly visible.} \label{fig:widthvsloc}
\end{figure}

Due to the complicated shape of the YSR spectra, we restrict the following analysis of the peaks to the peak position and full width at half maximum. These quantities provide direct insight into the nature of the YSR state-bulk interaction. Keeping in mind that the peak position directly depends on the strength of the exchange coupling (Fig.\ \ref{fig:shibacalc}), the extracted values are displayed in Fig.\ \ref{fig:widthvsloc}a) where the color coding represents the strength of the exchange coupling $J'$. The red line in Fig.\ \ref{fig:widthvsloc}a) shows twice the imaginary part of the order parameter of the first band $|2\Im\Delta_1(\omega)|$.

We observe a clear correlation between the broadening of the peak and the emergence of a finite $\Im\Delta_1(\omega)$ as calculated from Eq.\ \ref{eq:delta}. When the YSR peak position approaches the coherence peaks, where $|\Im\Delta_1(\omega)|$ increases abruptly, their width increases abruptly as well. Thus, there is a clear indication that the width of a YSR peak is related to the imaginary part of the superconducting order parameter.

\begin{figure}
\centerline{\includegraphics[width = 0.95\columnwidth]{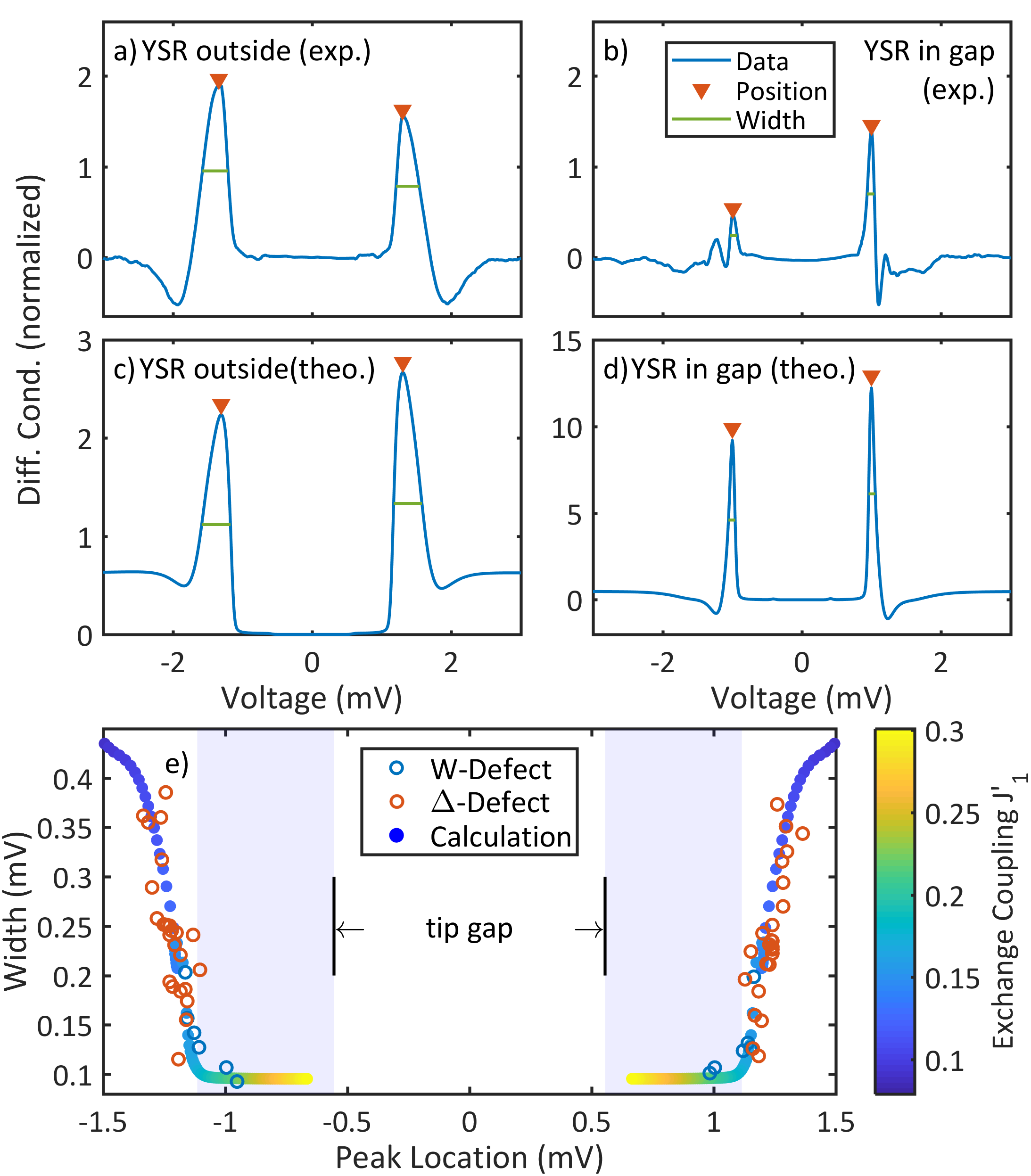}}
\caption{\textbf{Comparison of experiment with theory:} Peak positions and widths indicated for YSR states outside the gap in (a) and inside the gap in (b), where an unperturbed reference spectrum was subtracted. In (c) and (d), the corresponding calculated spectra are given. (e) Experimental YSR peak widths vs.\ peak position in comparison with the extracted YSR peak widths from the calculated differential conductance spectra. The W-shaped defect and the triangular defect ($\upDelta$) are color coded in blue and red, respectively. The blue shaded region indicates the sample gap.} \label{fig:expwidth}
\end{figure}

\subsection{Discussion}

In order to correlate the experimentally observed broadening of the YSR peak and its position with theory, we calculate the differential conductance from the density of states following Eq.\ \ref{eq:ysr}. The resulting spectra are shown in Fig.\ \ref{fig:widthvsloc}b) as a function of exchange coupling $J'_1$. The sharpening of the peaks when entering the region of the gap is clearly visible even though the width of the YSR peaks is now limited by the energy resolution of the STM \cite{ast_sensing_2016}. Two representative spectra of YSR states outside and inside the gap are plotted in Fig.\ \ref{fig:expwidth}a) and b), respectively. An unperturbed, averaged reference spectrum was subtracted to suppress the coherence peaks and isolate the YSR states. These experimental spectra can be compared to Fig.\ \ref{fig:expwidth}c) and d) showing two slices from Fig.\ \ref{fig:widthvsloc}b) with similar peak positions as in Fig.\ \ref{fig:expwidth}a) and b), respectively. The calculated spectra are normalized to the normal state conductance and scaled with the ratio $\eta$ (cf.\ Eq.\ \ref{eq:rho}). The subtraction of an unperturbed reference spectrum is not necessary in the calculated spectra, because the coherence peaks are inherently suppressed at the impurity centre where the DOS is calculated. Selecting theoretical spectra with matching peak positions, we observe good agreement between the corresponding panels concerning the width of the YSR peaks as well as their overall shape. We find, however, a reduced height in the experimental peaks, which we attribute to the measurement slightly above the (subsurface) impurity. Away from the scattering center of the impurity, intensity modulations can strongly reduce the peak height \cite{menard_coherent_2015}. We assume that they do not affect the peak position nor their width. Interestingly, the asymmetry of the YSR peaks is different in Fig.\ \ref{fig:expwidth}a) and b), which indicates a different Coulomb scattering potential $U'$ (assuming that the particle-hole asymmetry in the lattice Green's function does not change significantly between the impurities).

From each of the experimental and theoretical spectra, we extract the YSR peak widths and positions as indicated in Fig.\ \ref{fig:expwidth}a)-d) and display them in Fig.\ \ref{fig:expwidth}e). Excellent agreement is seen between the experimentally extracted values and the calculated spectra, indicating that there is indeed a strong correlation between the shape and position of the YSR states and the details of the underlying order parameter. Although W-defects seem to couple stronger than $\upDelta$-defects, both defects can be found in a range of exchange couplings (probably due to slightly different local environments), such that either defect can be used independently to infer the relation of the width to the imaginary part of the order parameter.

In a regular $s$-wave BCS superconductor the order parameter defines the position of the gap edge. YSR states must then always lie within the gap. Interband coupling in a multi-band superconductor leads to energy-dependent $\Delta_1$, which necessarily has a non-zero imaginary part \cite{toll_causality_1956}. The gap edge no longer corresponds to the value of $\Delta_1$ in this case and is generally found at lower energies. YSR states may then be found within the coherence peaks, but still at energies smaller than $\Delta_1$. We observe a marked increase in the line width of YSR states when they overlap with the coherence peaks.

The spectral functions on which these arguments are based result from the imaginary parts of respective Green's functions, which are commonly interpreted as single particle excitation spectra. The widths of the calculated features is determined by their relaxation rate into the ground state. The imaginary part of the order parameter can then be interpreted as a measure for the effective lifetime of YSR bound states. Indeed, an imaginary $\Delta_1$ renders the Green's function considered here non-Hermitian and indicates energy dissipation. These relaxation processes are strongest for YSR states located within the coherence peaks where the order parameter has a finite imaginary part. They sharpen up considerably within the superconducting gap, where the imaginary part of the order parameter tends to zero and there are few relaxation channels. The experimental linewidth is then limited by the resolution of our STM, which is determined by the interaction of the tunneling quasiparticles with the local electromagnetic environment according to the $P(E)$-description \cite{ast_sensing_2016,devoret_effect_1990,averin_incoherent_1990}.

This finite energy resolution of the STM obscures a direct observation of the peak width inside the superconducting gap. The full width at half maximum of the energy resolution function is dominated by the fluctuations of single charges at the junction capacitance \cite{ast_sensing_2016} and is usually much broader than the intrinsic peak width of just a few $\upmu$eV. Nevertheless, we surmise that even in the gap, there can be weak couplings to inelastic relaxation processes, which may be modeled effectively by a generic imaginary constant selfenergy, such as the Dynes parameter \cite{dynes_direct_1978}.

The correlation between YSR line width and $\Im\Delta_1(\omega)$ suggests that YSR states can also be exploited as local probes capable of measuring the imaginary part of a superconducting order parameter. A quantitative analysis would need to explicitly include the local modifications of the order parameter induced by the YSR states themselves \cite{salkola_spectral_1997,flatte_local_1997-1}. The nature and symmetry of the superconducting order parameter remains a subject of much debate across a wide field of materials in the scientific literature. Using YSR states as probes for the complex part of the superconducting order parameter could give valuable insight into the superconductivity of other non-trivial materials, especially those of an unconventional, proximity-induced, or topological nature.

\subsection{Conclusion}

In summary, we have shown that complex-valued energy dependent order parameters as they can result from multi-band superconductivity give rise to nontrivial interactions with local pair-breaking potentials. The resulting YSR states may not just exist in the superconducting gap, but can also overlap with the coherence peaks.  In the latter case, they acquire a substantial linewidth due to dissipative relaxation processes into the continuum. These processes are expressed in the order parameter, which can be interpreted as a ``rephrased'' self-energy. We show that the imaginary part of the order parameter is connected to lifetime broadening. Our analysis focused on the two-band $s$-wave superconductor NbSe$_2$ as a model system. We expect similar or even more complex interactions of YSR states with unconventional ($d$-wave, $p$-wave), proximity-induced, and/or topological superconductors. At the same time, we have demonstrated that YSR states may serve as sensitive probes for the imaginary part of the order parameter, which opens up new possibilities for understanding the intricacies of multi-band superconductivity as well as unconventional superconductors through the study of pair-breaking potentials.

\subsection{Acknowledgements}

We gratefully acknowledge stimulating discussions with Wolfgang Belzig, Hans Boschker, Carlos Cuevas, Reinhard Kremer, Bettina Lotsch, Jochen Mannhart, Fabian Pauly, and Markus Ternes. This work was funded in part by the ERC Consolidator Grant AbsoluteSpin (Grant No.\ 681164) and by the Center for Integrated Quantum Science and Technology (IQ$^\text{ST}$). CRV acknowledges funding from the COST Action ECOST-STSM-CA15128-010217-082276. SD acknowledges financial support from the Carl-Zeiss-Stiftung.

\onecolumngrid
\newpage
\begin{center}
\textbf{\large Supplementary Information}
\strut
\vspace{1em}
\end{center}
\setcounter{figure}{0}
\setcounter{table}{0}
\setcounter{equation}{0}
\renewcommand{\thefigure}{S\arabic{figure}}
\renewcommand{\thetable}{S\Roman{table}}
\renewcommand{\theequation}{S\arabic{equation}}
\twocolumngrid

\subsection{Tip and Sample Preparation}

The experiments were carried out in a scanning tunneling microscope (STM) operating at a base temperature of 15\,mK \cite{si_assig_10_2013}. The sample that was used was an 0.55\% Fe-doped NbSe$_2$ single crystal. To obtain a clean surface, the sample was cleaved with scotch tape in ultrahigh vacuum. The tip material was a polycrystalline vanadium wire, which was cut in air and prepared in ultrahigh vacuum by sputtering and field emission.

\subsection{F\MakeLowercase{e}-doped N\MakeLowercase{b}S\MakeLowercase{e}$_2$}

\begin{figure}
\centerline{\includegraphics[width = 0.75\columnwidth]{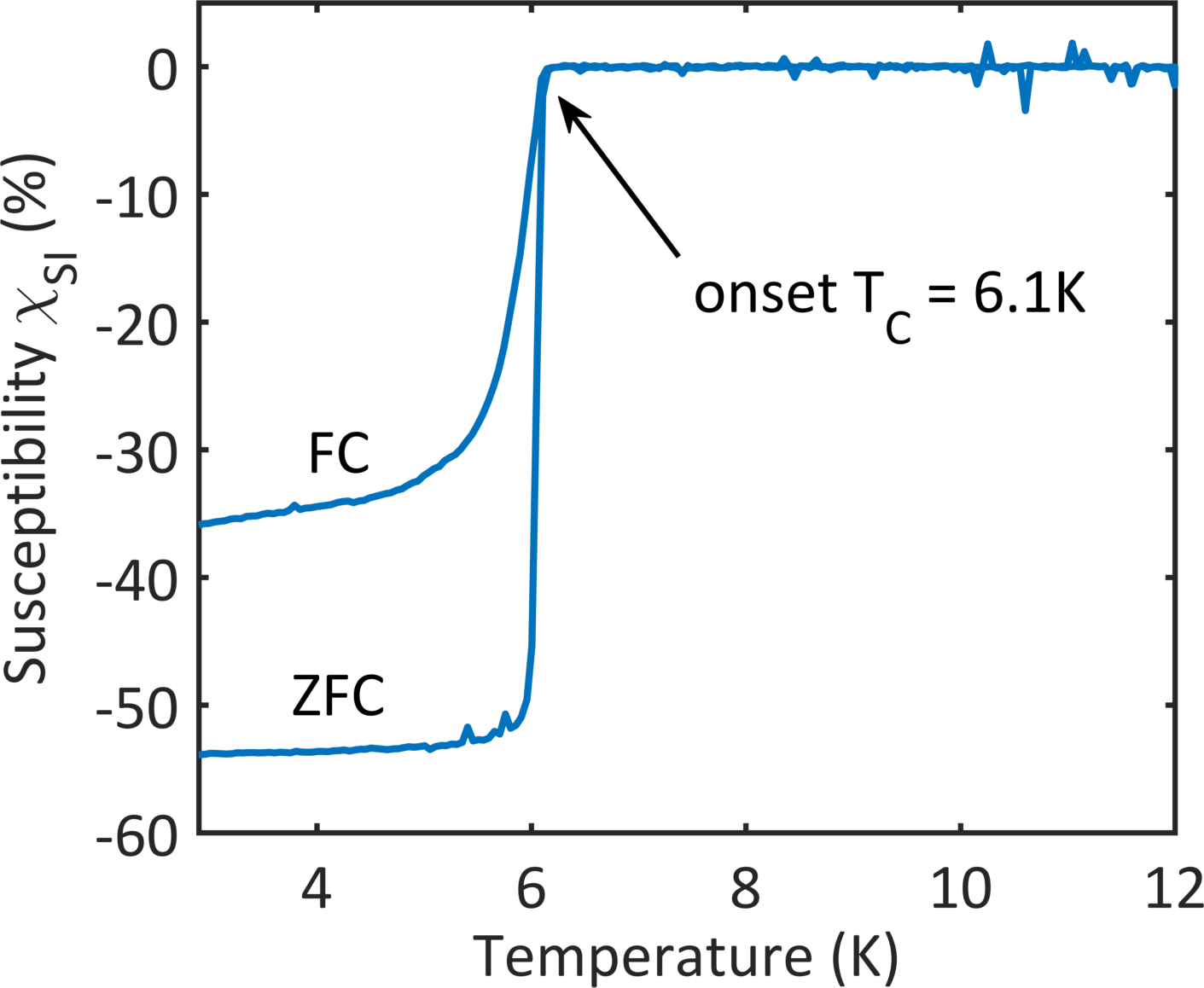}}
\caption{Zero-field cooled (ZFC) and field cooled (FC) susceptibility curves of the Fe-doped NbSe$_2$. The onset transition temperature is about 6.1\,K. The field cooled values are higher than the zero-field cooled values indicating a type-II superconductor.} \label{fig:squid}
\end{figure}

Single crystals of Fe-doped NbSe$_2$ were grown with chemical vapor transport. Powders of Nb and Fe in a ratio of 99.45:0.55 were mixed well and then placed in a quartz tube. Se chips were added in a stoichiometric ratio to yield Nb$_{0.9945}$Fe$_{0.055}$Se$_2$. Iodine was used as the transport agent. The sealed tube was heated to 900$^{\circ}$C with a temperature gradient of 75$^{\circ}$C for three weeks. The structure and composition of the crystals was confirmed with X-ray diffraction and EDX, respectively. The Fe content was too low to be detected in the EDX. The magnetic properties were measured on a MPMS (magnetic properties measurements system) from Quantum Design. From the MPMS measurements, we find an onset transition temperature $T_\text{C}=6.1\,$K (see Fig.\ \ref{fig:squid}).

\begin{figure}
\centerline{\includegraphics[width = 0.75\columnwidth]{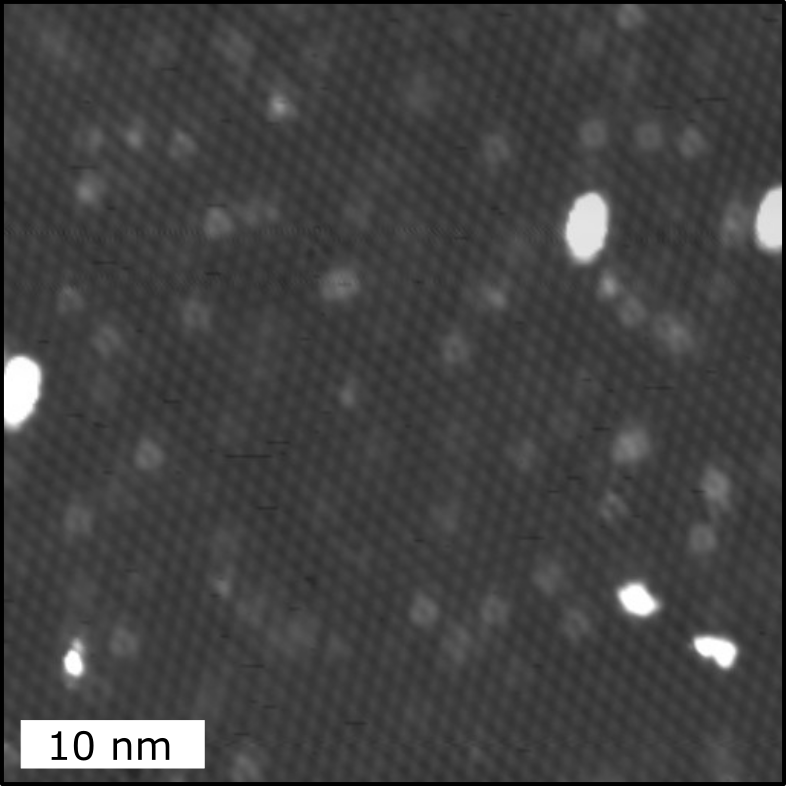}}
\caption{Topographic overview map of the Fe-doped NbSe$_2$ surface. A statistical distribution of subsurface Fe atoms is visible. The current setpoint was 20\,pA at a bias voltage of 100\,mV.} \label{fig:topo}
\end{figure}

A topographic overview map of the Fe-doped NbSe$_2$ surface can be seen in Fig.\ \ref{fig:topo}. A statistical distribution of subsurface Fe atoms is visible. The YSR spectra were taken at the geometric center of the defect. \\

\subsection{Calculating the differential Conductance}

The differential conductance $dI/dV$ was calculated from the tunneling current
\begin{equation}
I(V) = e\left(\vec{\mathit\Gamma}(V)-\cev{\mathit\Gamma}(V)\right),
\label{eq:iv}
\end{equation}
with the tunneling probability from tip to sample
\begin{widetext}
\begin{equation}
\vec{\mathit\Gamma}(V)=\frac{1}{e^2R_\text{T}}\int\limits^{\infty}_{-\infty}\int\limits^{\infty}_{-\infty}dEdE'\rho_\text{t}(E)\rho_\text{s}(E'+eV)
f(E)[1-f(E'+eV)]P(E-E').
\label{eq:tunprob}
\end{equation}
\end{widetext}
The other tunneling direction $\cev{\mathit\Gamma}(V)$ from sample to tip can be obtained by exchanging electrons and holes in Eq.\ \ref{eq:tunprob}. Here, $R_\text{T}$ is the tunneling resistance, $f(E)=1/(1+\exp(E/k_\text{B}T))$ is the Fermi function, and $\rho_\text{t}$, $\rho_\text{s}$ are the densities of states of tip and sample, respectively. The $P(E)$-function describes the exchange of energy with the environment during the tunneling process and is interpreted as the energy resolution function of the STM \cite{si_ast_sensing_2016}.

The parameter $\eta$ introduced in the main text weighs the density of states of the two bands in NbSe$_2$ as they appear in the experimental differential conductance spectra. It includes both the density of states for each band as well as the tunneling probability into each band. As it is not necessary to separate these contributions in this analysis, we have combined them with the density of states. This allows us to use Eq.\ \ref{eq:tunprob} with an overall tunneling resistance $R_\text{T}$, which can be directly determined from the experimental data.

\begin{figure}
\centerline{\includegraphics[width = 0.75\columnwidth]{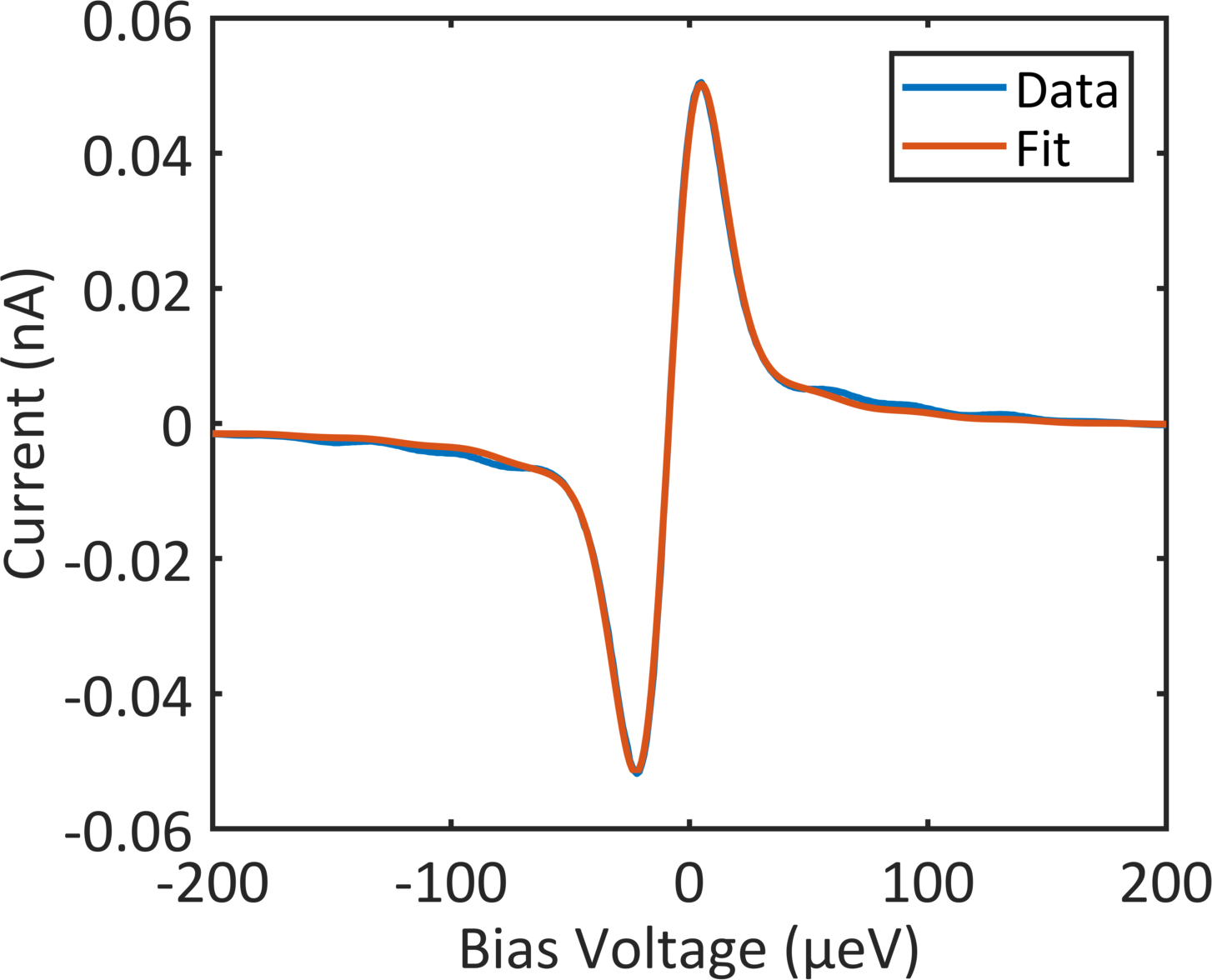}}
\caption{Current-voltage characteristics of the Josephson effect (blue curve) fitted with the $P(E)$-theory (red curve) to extract the parameters describing the $P(E)$-function. The spectrum was measured with a current setpoint of 15\,nA at a bias voltage of 4\,mV.} \label{fig:josfit}
\end{figure}

For the parameters in the $P(E)$-function, we perform an independent fit of a Josephson spectrum with the same macroscopic tip as used for the other spectra presented here \cite{si_jack_nanoscale_2015}. The Josephson effect in a scanning tunneling microscope at very low temperatures is governed by the dynamical Coulomb blockade and as such can be well described by the $P(E)$-theory \cite{si_devoret_effect_1990,si_averin_incoherent_1990,si_ingold_cooper-pair_1994}. The current-voltage characteristics $I(V)$ for the Josephson effect is given by
\begin{equation}
    I(V)=\frac{\pi e E^2_\text{J}}{\hbar}\left[P(2eV)-P(-2eV)\right],
\end{equation}
where $E_\text{J}$ is the Josephson coupling energy and $e$ is the electric charge. The experimental data and the corresponding fit are shown in Fig.\ \ref{fig:josfit}. The details of the $P(E)$-function pertaining to the tunnel junction of a scanning tunneling microscope are given elsewhere \cite{si_jack_nanoscale_2015}. The relevant fitting parameters for the $P(E)$-function are the tunnel junction capacitance $C_\text{J}=9.5$\,fF and an effective temperature $T=100\,$mK. The environmental impedance is modeled by a finite transmission line having an environmental resistance of $R_\text{env}=377\,\upOmega$, a tip resonance energy of $\hbar\omega_\text{res}=45\,\upmu$eV, and a resonance broadening factor $\alpha=0.7$. With these parameters, we can use the $P(E)$-function as the energy resolution function.

The density of states of the tip $\rho_\text{t}$ was modeled by the Maki equation \cite{si_Maki,si_Shiba} because of the intrinsic magnetic impurities in vanadium:
\begin{equation}
    \rho_\text{t}(\omega) = \Re\left[\frac{\omega+i\Gamma_\text{t}}{\sqrt{(\omega+i\Gamma_\text{t})^2-\Delta_\text{t}^2(\omega)}}\right],
\end{equation}
with
\begin{equation}
\Delta_\text{t}(\omega) = \Delta_\text{t}^\text{BCS} - \zeta_\text{t}\frac{\Delta_\text{t}(\omega)}{\sqrt{\Delta^2_\text{t}(\omega) - \omega^2}}.
\end{equation}
Here, $\Gamma_\text{t}$ is a phenomenological broadening term and $\zeta_\text{t}$ is a depairing parameter due to the interaction with a small concentration of magnetic impurities. For the vanadium tip, we find $\Delta_\text{t}^\text{BCS}=710\,\upmu$eV, $\Gamma_\text{t}=5\,\upmu$eV and $\zeta_\text{t}=33\,\upmu$eV. Note that due to the magnetic interaction modeled by the Maki equation, the tip gap ($2\times550\,\upmu$eV) is smaller than twice the order parameter of vanadium ($2\times710\,\upmu$eV).

Differential conductance spectra were recorded with a lock-in amplifier having a modulation amplitude of $20\,\upmu$V and a modulation frequency of $793\,$Hz. The resulting broadening is also included in the analysis and enters the calculated differential conductance spectra through an additional convolution.

\subsection{Momentum dependent vs.\ energy dependent order parameters}

At the heart of BCS theory, lies a potentially momentum dependent order parameter $\Delta_{\vek{k}}$, which appears in the momentum and energy dependent Green's function \cite{si_Schrieffer}:
\begin{equation}
    G(\omega,\vek{k}) = \frac{1}{\omega\sigma_0-\varepsilon_{\vek{k}}\sigma_3-\Delta_{\vek{k}}\sigma_1}
\end{equation}
Integrating over momentum space yields a momentum independent Green's function, where the functional dependence of the order parameter is projected onto an energy scale resulting in an energy dependent order parameter $\Delta(\omega)$ \cite{si_McMillan,si_Schrieffer},
\begin{equation}
    G(\omega) = -\pi\frac{\omega\sigma_0 - \Delta(\omega)\sigma_1}{\sqrt{\Delta^2(\omega) - \omega^2}},
\end{equation}
which has been used in the main text. In this sense, the momentum dependent order parameter $\Delta_{\vek{k}}$ is related to the energy dependent order parameter $\Delta(\omega)$ like the band structure $\varepsilon_{\vek{k}}$ is related to the density of states $n(\omega)$. As such, using an energy dependent order parameter does not ultimately exclude a momentum dependence, it even implies it. In our STM measurements, we are dealing with density of states measurements having no direct information of the momentum dependence. We, therefore, have chosen the simplest, suitable model to describe our experimental data since the extended complexity of employing a momentum dependent order parameter may change some details \cite{si_hayashi_effects_1997,si_rahn_gaps_2012,si_galvis_tilted_2017}, but not the general message of the manuscript.

\subsection{Numerical Solution for $\Delta_1(\omega)$ and $\Delta_2(\omega)$}

For a numerical solution at arbitrary energies, it is convenient to work with dimensionless quantities. We define
\begin{equation}
    u_1 = \frac{\omega}{\Delta_1(\omega)}\qquad u_2 = \frac{\omega}{\Delta_2(\omega)}.
\end{equation}
Using also the dimensionless abbreviations
\begin{equation}
    \tilde{\omega}=\frac{\omega}{\Delta_{1}^\text{BCS}};\ \tilde{\Gamma}_{12}=\frac{\Gamma_{12}}{\Delta_{1}^\text{BCS}};\  \tilde{\Gamma}_{21}=\frac{\Gamma_{21}}{\Delta_{1}^\text{BCS}};\ \nonumber
\end{equation}
\begin{equation}
    \tilde{\zeta}_{1}=\frac{\zeta_{1}}{\Delta_{1}^\text{BCS}};\
    \tilde{\zeta}_{2}=\frac{\zeta_{2}}{\Delta_{1}^\text{BCS}};\ \delta=\frac{\Delta_{2}^\text{BCS}}{\Delta_{1}^\text{BCS}},
\end{equation}
Eq.\ (1) of the main text transforms into
\begin{eqnarray}
    \tilde{\omega} &=& u_1 - \tilde{\Gamma}_{12}\frac{u_2 - u_1}{\sqrt{1-u^2_2}} - \tilde{\zeta}_1\frac{u_1}{\sqrt{1-u^2_1}}\\
    \tilde{\omega} &=& u_2\delta - \tilde{\Gamma}_{21}\frac{u_1 - u_2}{\sqrt{1-u^2_1}} - \tilde{\zeta}_2\frac{u_2}{\sqrt{1-u^2_2}}.
\end{eqnarray}
We have referenced all variables to $\Delta_{1}^\text{BCS}$, because if we were to reference each equation to its own order parameter, the problem becomes ill-defined, if $\Delta_{2}^\text{BCS}=0$. Avoiding this problem allows us to work with dimensionless quantities.
For the Newton-Raphson method, we have to cast the above equations into a different form defining
\begin{equation}
    y_1 = \frac{u_1}{\sqrt{1-u^2_1}};\ y_2 = \frac{1}{\sqrt{1-u^2_1}};\ y_3 = \frac{u_2}{\sqrt{1-u^2_2}};\ y_4 = \frac{1}{\sqrt{1-u^2_2}}
\end{equation}
We find
\begin{eqnarray}
    y_1        - \tilde{\Gamma}_{12}(y_3y_2 - y_1y_4) - \tilde{\zeta}_1y_1y_2 - \tilde{\omega}y_2 & = & 0 \nonumber\\
    \delta y_3 - \tilde{\Gamma}_{21}(y_1y_4 - y_3y_2) - \tilde{\zeta}_2y_3y_4 - \tilde{\omega}y_4 & = & 0 \nonumber\\
    y^2_2 - y^2_1 - 1 & = & 0 \nonumber\\
    y^2_4 - y^2_3 - 1 & = & 0 \label{eq:newton}
\end{eqnarray}
These equations can be easily solved using a multi-dimensional Newton-Raphson method \cite{si_Worledge,si_Newton}.

To solve the system of equations in Eq.\ \ref{eq:newton}, we define a multi-valued function $f(\vec{y})$, whose input is a four-dimensional vector $\vec{y}=(y_1, y_2, y_3, y_4)$ and whose output are the values of the four quantities on the left hand sides in Eq.\ \ref{eq:newton}. With the Newton-Raphson method, we seek a physical solution to the equation
\begin{equation}
    f(\vec{y})=0.
\end{equation}
This is done iteratively by calculating the next value from the equation
\begin{equation}
    f(\vec{y}_n) + J(\vec{y}_n)\cdot\Delta\vec{y}_n = 0\ \Rightarrow\ \Delta\vec{y}_n = -J(\vec{y}_n)^{-1}f(\vec{y}_n)
\end{equation}
where $J(\vec{y})$ is the Jacobian matrix of $f(\vec{y})$ with
\begin{equation}
    J(\vec{y})_{ij} = \frac{\partial f_i}{\partial y_j}
\end{equation}
The next iteration value is then
\begin{equation}
    \vec{y}_{n+1} = \vec{y}_{n} + \Delta\vec{y}_{n}.
\end{equation}
When calculating the values $\vec{y}$ as function of energy $\omega$, the best starting point is $\omega = 0$ and then to use the previous value as the starting point for the next value of $\omega$. A universal starting point for $\omega=0$ is $\vec{y} = (0, 1, 0, 1)$, which works well for a broad set of parameters.

\subsection{Analysis of the coupled two-band superconductivity}

We start from McMillan's coupled equations for a two-band superconductor with bare (energy independent and real-valued) BCS gaps $\Delta_{1}^{\rm BCS}\equiv \Delta_0\neq 0 $ and $\Delta_{2}^{\rm BCS}=0$, respectively, i.e.
\begin{eqnarray}
\Delta_1(\omega)&=&\Delta_0-\Gamma_{12} \frac{\Delta_1(\omega)-\Delta_2(\omega)}{\sqrt{\Delta_2^2-\omega^2}}\\
\Delta_2(\omega)&=&-\Gamma_{21} \frac{\Delta_2(\omega)-\Delta_1(\omega)}{\sqrt{\Delta_1^2-\omega^2}}\,.
\label{mcmillan}
\end{eqnarray}
Here $\Gamma_{12}, \Gamma_{21}$ denote interband coupling rates. The second equation can be easily solved to read
\begin{equation}
\Delta_2=\frac{\Gamma_{21}\, \Delta_1}{\Gamma_{21}+\sqrt{\Delta_1^2-\omega^2}}\,
\label{deltaa}
\end{equation}
which means that a complex-valued $\Delta_1$ implies a complex-valued $\Delta_2$ and vice versa.

Now, in case of NbSe$_2$ one may approximately assume $\Gamma_{12}\ll \Gamma_{21}\sim \Delta_0$ which means that $\Delta_0 \sim \Delta_1 \sim \Delta_2$ as long as only orders of magnitude are concerned. Accordingly, one writes $\Delta_1=\Delta_0+\delta_1$ with $|\delta_1|\sim \Gamma_{12}\ll \Delta_0$ and solves the first equation in (\ref{mcmillan}) for $\delta_1$, i.e.,
\begin{equation}
\delta_1=-\Gamma_{12} \frac{\Delta_0-\Delta_2}{\Gamma_{12}+\sqrt{\Delta_2^2-\omega^2}}
\end{equation}
while for (\ref{deltaa}) one obtains by putting $\Delta_1\approx \Delta_0$
\begin{equation}
\Delta_2=\frac{\Gamma_{21}\, \Delta_0}{\Gamma_{21}+\sqrt{\Delta_0^2-\omega^2}}\, .
\label{delta2}
\end{equation}
Using this latter result, one finds the explicit expression for the correction to $\Delta_1$ to read
\begin{equation}
\delta_1=-\frac{\Delta_0\Gamma_{12}\, \sqrt{\Delta_0^2-\omega^2}}{\Gamma_{12}(\Gamma_{21}+\sqrt{\Delta_0^2-\omega^2})+\sqrt{\Gamma_{21}^2 \Delta_0^2-\omega^2\, (\Gamma_{21}+\sqrt{\Delta_0^2-\omega^2})^2}}\, .
\label{delta1}
\end{equation}

One can now distinguish three regions in frequency space:
\begin{alexdes}

\item[1.\ |\omega|<\omega_c]\strut\\
    Here, $\omega_c=\alpha\, \Delta_0$, where $\alpha<1$ is the positive solution of $\alpha^3+\alpha^2+ \epsilon^2 \, \alpha -\epsilon^2=0$ with $\epsilon=\Gamma_{21}/\Delta_0\approx \sqrt{2}$. The only real solution is $\alpha\approx 0.65$. In this regime, $\Delta_1$, $\Delta_2$ are both real-valued with $\Delta_1, \Delta_2>0$. The energy $\omega_c $ defines the threshold, below which imaginary parts in both order parameters are absent. Thus $\omega_c$ determines the width of the superconducting gap. Note that the gap widths for both bands are identical as an imaginary part in one order parameter induces an imaginary part in the other one and vice versa.

\item[2.\ \omega_c<|\omega|\lesssim\Delta_0]\strut\\
    In this regime, $\delta_1$ is complex-valued so that also $\Delta_2$ acquires a small imaginary part. The root is taken according to $\sqrt{\omega_c -\omega}\to -i \sqrt{\omega-\omega_c}$.  Accordingly, Im$\{\delta_1(\omega>0)\}<0$ so that Im$\{\Delta_2(\omega>0)\}<0$. In particular, $\Delta_2(\omega_c)\approx -0.65\, \Delta_0$ while $\delta_1(\omega_c)/\Delta_0=-\sqrt{\Delta_0^2-\omega_c^2}/[\Gamma_{12}(\Gamma_{21}+\sqrt{\Delta_0^2-\omega_c^2})]\approx - 0.349\, \Delta_0$. Note that this perturbative treatment is less accurate near the boundaries of their respective frequency domains.

\item[3.\ \Delta_0\lesssim |\omega|]\strut\\
    In this regime, ${\rm Im}\{\delta_1\}$ and $\Delta_2$ approach zero and ${\rm Im}\{\Delta_2(\omega>0)\}>0$ meaning that  ${\rm Im}\{\Delta_2\}=0$ at a frequency near $\Delta_0$.

\end{alexdes}

The above treatment can be refined by replacing $\Delta_0\to \Delta_0+\delta_1$ in (\ref{delta2}) and using the expression (\ref{delta1}). One then finds $\Delta_2(\omega_c)\approx \Delta_1(\omega_c) $ in agreement with the full numerical solution (see Fig.\ \ref{fig:fitsops}).

\end{document}